\documentclass[preprint]{elsart}

\newcommand{\epl}{Europhys. Lett.\ }

\newcommand{\pr}{Phys. Rev.\ }
\newcommand{\pra}{Phys. Rev. A\ }
\newcommand{\prl}{Phys. Rev. Lett.\ }

\newcommand{\oc}{Opt. Commun.\ }

\usepackage{graphicx}% eps files
\usepackage{bm}% bold math
\usepackage{amsmath}% equation formatting

\begin{document}

\begin{frontmatter}

\title{Phase-space analysis of bosonic spontaneous emission}

\author[UQ]{M.~K. Olsen},
\author[Ulm]{L.~I. Plimak},
\author[Camerino]{S. Rebi\'c} and
\author[UQ]{A.S. Bradley}

\address[UQ]{ARC Centre of Excellence for Quantum-Atom Optics, 
School of Physical Sciences, University of Queensland, Brisbane, 
Qld 4072, Australia} 
\address[Ulm]{Universit\"at Ulm,
Abteilung f\"ur Quantenphysik,
D-89069 Ulm,
Germany}
\address[Camerino]{INFM and Dipartimento di Fisica, Universit\`a di Camerino, I-62032 Camerino (MC), Italy}

\begin{abstract}    

We present phase-space techniques for the modelling of spontaneous emission in two-level bosonic atoms. The positive-P representation is shown to give a 
full and complete description and can be further developed to give exact treatments of the interaction of degenerate bosons with the electromagnetic field in a given experimental situation. The Wigner representation, even when truncated at second order, is shown to need a doubling of the 
phase-space to allow for a positive-definite diffusion matrix in the appropriate Fokker-Planck equation and still fails to agree with the full quantum results of the positive-P representation. We show that quantum statistics and correlations between the ground and excited states affect the dynamics of the emission process, so that it is in general non-exponential.

PACS numbers: 42.50.Dv,03.65.Yz,03.75.-b
\end{abstract}

\begin{keyword}
Spontaneous emission, phase-space representations, many-body effects.
\end{keyword}

\end{frontmatter}

\section{Introduction}
\label{subsec:intro}

The study of spontaneous emission has a long history, beginning with the famous Einstein A and B coefficients~\cite{Einstein} and the Wigner-Weisskopf law~\cite{WW}. A comprehensive 
account can be found in Agarwal~\cite{Agarwal}. Previous studies have considered atoms which are not within one de Broglie wavelength of each other, so that the fermionic or bosonic nature of the atoms need not be considered. This condition holds for dilute high-temperature atomic samples, but not for degenerate quantum gases. In particular, it does not hold for Bose-Einstein condensates~\cite{Scott} and now that these are readily available in laboratories around the world, it is of some interest to develop methods which will 
allow for a full quantum treatment of both the atomic and electromagnetic fields.
In this paper we develop and demonstrate phase-space methods, using the positive-P~\cite{P+} and 
Wigner~\cite{Wigner} representations, for the treatment of spontaneous 
emission from two-level bosonic atoms. We use a simple model without spatial dependence of the atomic fields and without collisional and dipole-dipole interactions, although these can readily be included for larger atomic samples, along with other details which would be necessary to model a realistic experimental situation. The simplicity of our model allows us to consider different quantum states of the atomic fields and demonstrate clearly that correlations which can build up between the ground and excited states have a noticeable effect on the dynamics. We can also give a clear demonstration of the difference between the quantum mechanical predictions of the positive-P representation and an approximation which we derive from the Wigner representation.

\section{Formalism}
\label{sec:formalismo}

To begin, we define $\psi(x)$ as a wavefunction for a bosonic atomic field, with
\begin{equation}
\psi(x)=\sum_{j}a_{j}\psi_{j}(x),
\label{eq:psidiff}
\end{equation}
the $\psi_{j}(x)$ are then wave-functions for different (distinguishable) atomic fields.
In second quantisation, the expansion coefficients 
are changed to operators 
so that $\hat{a}_{j}^{\dag}\hat{a}_{j}$ becomes the number operator for atoms of the $j$th type. 
For simplicity, we consider one chemical species of atom  
with two electronic levels, $j=a,b$, for the ground and excited states, respectively, and set $\hat{a}_{a}=\hat{a}$ and $\hat{a}_{b}=\hat{b}$. 
Having introduced the second-quantised picture we could also include 
fermionic atoms, for which Eq.~\ref{eq:psidiff} is not 
defined, and use fermionic phase-space methods which are under development~\cite{Joel}. However, in this work we will consider only bosonic atoms. Since the total number of atoms is conserved, the 
physically relevant (bosonic) operator combinations can be defined by
\begin{eqnarray}
\hat{a}^{\dag}\hat{b}|n_{a},n_{b}\rangle &=& \sqrt{(n_{a}+1)n_{b}}|n_{a}+1,n_{b}-1\rangle,\nonumber\\
\hat{b}^{\dag}\hat{a}|n_{a},n_{b}\rangle &=& \sqrt{(n_{b}+1)n_{a}}|n_{a}-1,n_{b}+1\rangle,
\label{eq:opdiff}
\end{eqnarray}
where 
$|n_{a},n_{b}\rangle$ signifies an atomic field with $n_{a}$ atoms in the ground state 
and $n_{b}$ atoms in the excited state.

We will consider first the case of a single-atom Fock state, where $\langle\hat{a}^{\dag}\hat{a} + \hat{b}^{\dag}\hat{b}\rangle = 1$. 
The atomic Fock space then reduces to two dimensions, with basis vectors 
$|0,1\rangle$ and $|1,0\rangle$. 
The combinations of operators (\ref{eq:opdiff}) then play an analogous role to the normal 
spin raising and lowering operators.
We can now make the following correspondences with the normal Pauli spin operators:
\begin{equation}
\sigma^{+} \leftrightarrow \hat{b}^{\dag}\hat{a},\:\:\:\:\sigma^{-} \leftrightarrow \hat{a}^{\dag}\hat{b},\:\:\:\:
\sigma_{z} \leftrightarrow \frac{1}{2}\left(\hat{b}^{\dag}\hat{b}-\hat{a}^{\dag}\hat{a}\right),
\label{eq:pauli} 
\end{equation}
and look at the commutation relations. We find
\begin{eqnarray}
& &\left[\hat{b}^{\dag}\hat{a},\hat{a}^{\dag}\hat{b}\right] = \hat{b}^{\dag}\hat{b}-\hat{a}^{\dag}\hat{a},\nonumber\\
& &\left[\hat{b}^{\dag}\hat{a},\hat{b}^{\dag}\hat{b}-\hat{a}^{\dag}\hat{a}\right] = -2\hat{b}^{\dag}\hat{a},\nonumber\\
& &\left[\hat{a}^{\dag}\hat{b},\hat{b}^{\dag}\hat{b}-\hat{a}^{\dag}\hat{a}\right] = 2\hat{a}^{\dag}\hat{b},
\label{eq:commutators}
\end{eqnarray}
all completely equivalent to those of two-level atomic operators such  
as used in the well-known Jaynes-Cummings model~\cite{JCoriginal}, as long as we are considering a single atom. We note here that a similar formalism has previously been used by Bonifacio and 
Preparata~\cite{Italianos}, although they did not develop a phase-space representation of the problem as we will do in what follows.

Not unexpectedly, this simple relation to the operators used in the Jaynes-Cummings model 
does not survive beyond the single-atom Fock-state case. 
Consider for example the expressions for the probability of an atom being in a particular state. 
In the latter,  
(with $P_{jj}$ signifying the probability of being in state $j$)
\begin{equation}
P_{bb} = \langle \sigma^{+}\sigma^{-}\rangle,\:\:\:\: P_{aa} = \langle \sigma^{-}\sigma^{+}\rangle, 
\label{eq:paulipops}
\end{equation}
and, if we naively use these correspondences (\ref{eq:pauli}), we find
\begin{eqnarray}
\langle \sigma^{+}\sigma^{-}\rangle &\leftrightarrow& \langle \hat{b}^{\dag}\hat{a}\hat{a}^{\dag}\hat{b}\rangle = \langle\hat{b}^{\dag}\hat{b}\rangle 
+\langle\hat{b}^{\dag}\hat{b}\hat{a}^{\dag}\hat{a}\rangle,\nonumber\\
\langle \sigma^{-}\sigma^{+}\rangle &\leftrightarrow& \langle \hat{a}^{\dag}\hat{b}\hat{b}^{\dag}\hat{a}\rangle = \langle\hat{a}^{\dag}\hat{a}\rangle 
+\langle\hat{b}^{\dag}\hat{b}\hat{a}^{\dag}\hat{a}\rangle,
\label{eq:bosonpops}
\end{eqnarray}
which are obviously not the correct probabilities. This is because the Pauli operators describe a single atom with one fermion (electron) which can be in either the ground or excited state. This means that the density matrix need only be $2\times 2$, whereas we wish to investigate bosonic atoms where the density matrix is, in priciple, infinite. 
To solve this problem, we define
\begin{equation}
P_{bb} = \frac{\langle \hat{b}^{\dag}\hat{b}\rangle}{\langle\hat{b}^{\dag}\hat{b}\rangle+\langle\hat{a}^{\dag}\hat{a}\rangle},\:\:\:\:
P_{aa} = \frac{\langle \hat{a}^{\dag}\hat{a}\rangle}{\langle\hat{b}^{\dag}\hat{b}\rangle+\langle\hat{a}^{\dag}\hat{a}\rangle},
\label{eq:truepops}
\end{equation}
which are now the correct probabilities. 
We may define the atomic coherences as
\begin{equation}
P_{ba} = \langle\sigma^{-}\rangle \leftrightarrow \langle \hat{a}^{\dag}\hat{b}\rangle,\:\:\:\:
P_{ab} = \langle\sigma^{+}\rangle \leftrightarrow \langle \hat{b}^{\dag}\hat{a}\rangle.
\label{coherences}
\end{equation}
We may now write any master equation in terms of these bosonic operators and use the standard mappings~\cite{Crispin} to find Fokker-Planck and 
stochastic differential equations in the relevant phase-space representations. As an example to demonstrate our method, we will examine spontaneous emission from excited bosonic atoms.

\section{Spontaneous decay}
\label{sec:emit}

Spontaneous decay into a zero temperature thermal bath can be modelled by the Hamiltonian
\begin{equation}
H_{bath}=\hbar\left(\hat{a}\hat{b}^{\dag}\Gamma+\hat{a}^{\dag}\hat{b}\Gamma^{\dag}\right),
\label{eq:banho}
\end{equation}
where the $\Gamma$ are operators for the bath quanta.
Following the usual methods~\cite{Danbook}, this leads to the master equation
\begin{equation}
\frac{d\rho}{dt} = \frac{\kappa}{2}\left(2\hat{a}^{\dag}\hat{b}\rho\hat{a}\hat{b}^{\dag}-\hat{a}\hat{b}^{\dag}\hat{a}^{\dag}\hat{b}\rho-
\rho\hat{a}\hat{b}^{\dag}\hat{a}^{\dag}\hat{b}\right).
\label{eq:mestra}
\end{equation}

Looking at Eq.~\ref{eq:mestra}, we can immediately 
find equations for expectation values as 
\begin{equation}
\frac{d\langle\hat{x}\rangle}{dt}=Tr\left\{\hat{x}\frac{d\rho}{dt}\right\}.
\label{eq:esperado}
\end{equation}    
We find 
\begin{eqnarray}
\frac{d\langle\hat{b}^{\dag}\hat{b}\rangle}{dt} &=& -\kappa\left(\langle\hat{b}^{\dag}\hat{b}\rangle+\langle\hat{a}^{\dag}\hat{a}\hat{b}^{\dag}\hat{b}
\rangle\right),\nonumber\\
\frac{d\langle\hat{a}^{\dag}\hat{a}\rangle}{dt} &=& \kappa\left(\langle\hat{b}^{\dag}\hat{b}\rangle+\langle\hat{a}^{\dag}\hat{a}\hat{b}^{\dag}\hat{b}
\rangle\right),
\label{eq:decaimentomestra}
\end{eqnarray}
so that any correlations which build up between the two atomic fields can be important in the dynamics of the decay. These may be expected to depend on the initial atomic quantum states, something which is easily modelled using positive-P and Wigner representations. In the standard techniques which use 
a density matrix approach, approximations are usually necessary to model anything but number states, which are not the only possible choice when we wish to treat degenerate quantum gases.

The choice of Hamiltonian (\ref{eq:banho}) shows clearly that our analysis will {\em not\/} describe superrradiant emission. The latter 
may be defined as a feedback of the 
radiated field on atoms, whereas, by using a bath for the 
emitted field this feedback is formally eliminated. Irrespective of how 
important superrradiation may happen to be 
in a real experiment, the effects we describe  
in this paper are rooted in the many-body atom-field statistics.

\section{Positive-P representation}
 
We will first develop our stochastic equations in the positive-P representation.
Using the usual techniques~\cite{Crispin}, the master equation (\ref{eq:mestra}) is mapped onto a Fokker-Planck equation for the 
P-function~\cite{Glauber,Sud},
\begin{eqnarray}
\frac{dP}{dt} = & & \left\{-\frac{\kappa}{2}\left[\frac{\partial}{\partial\alpha}|\beta|^{2}\alpha+\frac{\partial}{\partial\alpha^{\ast}}|\beta|^{2}\alpha^{\ast}
-\frac{\partial}{\partial\beta}(|\alpha|^{2}+1)\beta-\frac{\partial}{\partial\beta^{\ast}}(|\alpha|^{2}+1)\beta^{\ast}\right]
\right.\nonumber\\
& &\left.
+\frac{\kappa}{2}\left[\frac{\partial^{2}}{\partial\alpha\partial\alpha^{\ast}}2|\beta|^{2}-\frac{\partial^{2}}{\partial\alpha\partial\beta}\alpha\beta-
\frac{\partial^{2}}{\partial\alpha^{\ast}\partial\beta^{\ast}}\alpha^{\ast}\beta^{\ast}\right]\right\}P(\alpha,\beta,t).
\label{eq:maisFPE}
\end{eqnarray}
The diffusion matrix, $D$, of Eq.~\ref{eq:maisFPE}, is not positive-definite so we must use the positive-P representation. The matrix can be factorised as
\begin{equation}
B = \sqrt{\kappa}\left[\begin{array}{cccccc}
\frac{i}{2}\sqrt{\alpha\beta} & 0 & -\frac{1}{2}\sqrt{\alpha\beta} & 0 & \sqrt{\frac{\beta^{+}\beta}{2}} & i\sqrt{\frac{\beta^{+}\beta}{2}}\\
0 & \frac{i}{2}\sqrt{\alpha^{+}\beta^{+}} & 0 & -\frac{1}{2}\sqrt{\alpha^{+}\beta^{+}} & \sqrt{\frac{\beta^{+}\beta}{2}} & -i\sqrt{\frac{\beta^{+}\beta}{2}}\\
\frac{i}{2}\sqrt{\alpha\beta} & 0 & \frac{1}{2}\sqrt{\alpha\beta} & 0 & 0 & 0\\
0 & \frac{i}{2}\sqrt{\alpha^{+}\beta^{+}} & 0 & \frac{1}{2}\sqrt{\alpha^{+}\beta^{+}} & 0 & 0\end{array}\right],
\label{eq:matrizdemerda}
\end{equation}
where $D=BB^{\text{T}}$.
This immediately allows us to write the It\^o stochastic equations
\begin{eqnarray}
\frac{d\alpha}{dt} &=& \frac{\kappa}{2}\beta^{+}\beta\alpha-\frac{1}{2}\sqrt{\kappa\alpha\beta}\left(\eta_{3}-i\eta_{1}\right)+
\sqrt{\frac{\kappa\beta^{+}\beta}{2}}\left(\eta_{5}+i\eta_{6}\right),\nonumber\\
\frac{d\alpha^{+}}{dt} &=& \frac{\kappa}{2}\beta^{+}\beta\alpha^{+}-\frac{1}{2}\sqrt{\kappa\alpha^{+}\beta^{+}}\left(\eta_{4}-i\eta_{2}\right)+
\sqrt{\frac{\kappa\beta^{+}\beta}{2}}\left(\eta_{5}-i\eta_{6}\right),\nonumber\\
\frac{d\beta}{dt} &=& -\frac{\kappa}{2}\left(\alpha^{+}\alpha+1\right)\beta+\frac{1}{2}\sqrt{\kappa\alpha\beta}\left(\eta_{3}+i\eta_{1}\right),\nonumber\\
\frac{d\beta^{+}}{dt} &=& -\frac{\kappa}{2}\left(\alpha^{+}\alpha+1\right)\beta^{+}+\frac{1}{2}\sqrt{\kappa\alpha^{+}\beta^{+}}\left(\eta_{4}+i\eta_{2}\right),
\label{eq:decaimento}
\end{eqnarray}
which may be numerically integrated. Note that, as always in the positive-P representation, there is a correspondence between the operators $\hat{a},\hat{b},\hat{a}^{\dag},\hat{b}^{\dag}$ and the c-number variables $\alpha,\beta,\alpha^{+},\beta^{+}$, such that
\begin{equation}
\overline{(\alpha^{+})^{m}(\beta^{+})^{n}
\alpha^{p}\beta^{q}}\rightarrow\langle (\hat{a}^
{\dag})^{m}(\hat{b}^{\dag})^{n}\hat{a}^{p}
\hat{b}^{q}\rangle,
\label{eq:classquant}
\end{equation}
where the left hand side is a classical average over stochastic trajectories and the right hand side is a quantum-mechanical expectation value.

Before we resort to stochastic integration, we will see what information we can get from the It\^o equations about the average decay rates, with the caveat that a mean-field approach 
may not be particularly meaningful for small atomic samples. Using the rules of It\^o calculus, where
\begin{equation}
d(xy)=y\,dx+x\,dy+dy\,dx,
\label{eq:Itorules}
\end{equation} 
we find mean-field equations for the 
populations in the ground ($N_{a}=\overline{\alpha^{+}\alpha}$) and excited ($N_{b}=\overline{\beta^{+}\beta}$) states, 
\begin{eqnarray}
\frac{dN_{a}}{dt} &=& \kappa\left(N_{a}+1\right)N_{b},\nonumber\\
\frac{dN_{b}}{dt} &=& -\kappa\left(N_{a}+1\right)N_{b}.
\label{eq:Itomeans}
\end{eqnarray}
For those not familiar with stochastic calculus, we note that merely dropping the noise terms from Eq.~\ref{eq:decaimento} would not give the correct equations for the populations, although this would be the correct procedure if we were using Stratonovich equations.
With $N_{T}$ the total number of atoms, we may now write an equation for $N_{b}$,
\begin{eqnarray}
\frac{dN_{b}}{dt} &=& -\kappa\left(N_{a}+1\right)N_{b}\nonumber\\
&=& -\kappa\left(N_{T}-N_{b}+1\right)N_{b},
\label{eq:dNbdt}
\end{eqnarray}
which may be solved to give 
\begin{equation}
N_{b}(t)=\frac{N_{b}(0)\left(N_{T}+1\right)}{N_{b}(0)\left[N_{T}+1-N_{b}(0)\right]\mbox{e}^{(N_{T}+1)\kappa t}}.
\label{eq:Nbsoldet}
\end{equation}
In the case of one atom initially excited, this simplifies to
\begin{equation}
N_{b}(t) = \mbox{sech}(\kappa t)\mbox{e}^{-\kappa t},
\label{eq:sechsol}
\end{equation}
although we would not expect mean-field solutions to be accurate for such small numbers of quanta.
However, we do see that the decay 
rate is proportional to the number of quanta which can be in the final state, 
in a manner reminiscent of Fermi's golden rule~\cite{Sakurai}. A work on superradiance, by Rehler and Eberly~\cite{Rehler}, gives an expression for the atomic energy (their Eq.~5.1), which, in the limit that all atoms 
are within a wavelength of each other so that their parameter $\mu$ is equal to $1$, becomes
\begin{equation}
W(t) = -\frac{1}{2}N_{b}(0)\left[\mbox{e}^{(N_{b}(0)+1)\kappa t}-(N_{b}(0)+2)\right]/\left(\mbox{e}^{(N_{b}(0)+1)\kappa t}+N_{b}(0)\right).
\label{eq:Eberly}
\end{equation}
We note that, once we redefine their energy scale so that the energy of the ground state is zero rather than negative and all atoms are initially excited, this gives the same result as our Eq.~\ref{eq:Nbsoldet}. (In Ref.~\cite{Rehler},
the energy of a ground (excited) state atom is defined as $-(+)\hbar\omega/2$.)
It is interesting 
that the same result has been arrived at by different means, although our result allows for some initial population in the ground state and, as shown below by the 
numerical results, is only an approximation to the real result, which we will show to depend on the quantum state of the atomic ensemble. In the regime where all atoms were within an optical wavelength, we would also expect 
dipole-dipole forces and maybe even collisional processes to become important. We stress here that the non-exponential decay predicted by our approach is not the classical superradiance, but rather is due to correlations between the ground and excited state atomic fields.

\begin{figure}
\begin{center} 
\includegraphics[width=0.9\columnwidth]{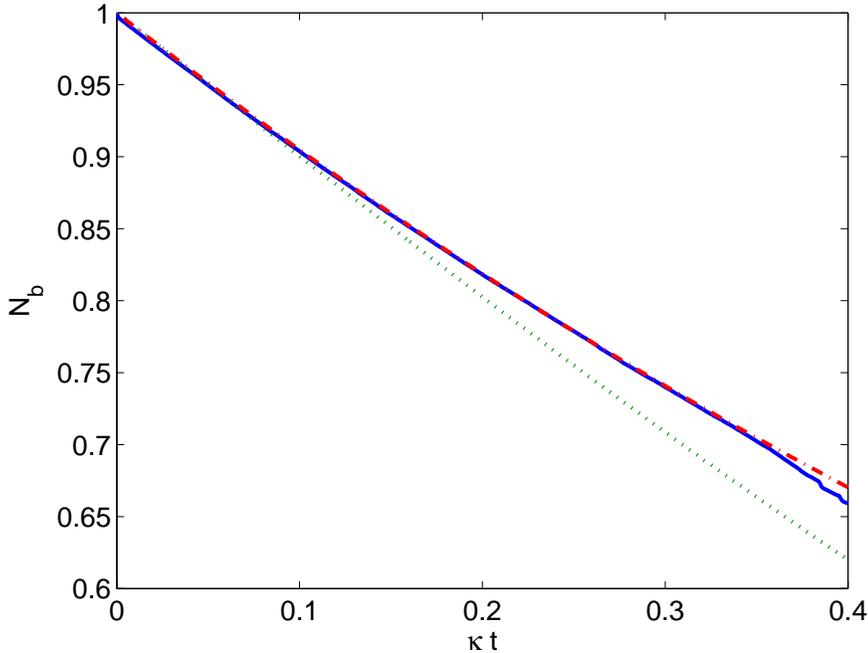}
\end{center} 
\caption{Positive-P solution averaged over $3.4\times 10^{5}$ stochastic trajectories (solid line) for decay of an $N=1$ Fock state, with $\kappa=0.4$. The exponential solution (dash-dotted line) and the mean-field analytical solution (dotted line) are shown for comparison.}
\label{fig:Fock}
\end{figure}

To obtain the full quantum results, the Stratonovich version of Eq.~\ref{eq:decaimento} was numerically integrated. As usual in quantum stochastic integration, we must decide on which initial conditions to use for 
our atomic fields. The simplest initial condition to model in the positive-P representation is a coherent state, $|\alpha_{0}\rangle$, which can be represented by the pseudoprobability distribution $P(\alpha,\alpha^{+})=\delta(\alpha-\alpha_{0})\delta(\alpha^{+}-\alpha_{0}^{\ast})$. Other 
quantum states which may arise naturally when we treat atoms are the chaotic state and the Fock state, which has a fixed number of atoms. It has been shown that any quantum state may be represented by the positive-P distribution~\cite{P+}
\begin{equation}
P(\alpha,\alpha^{+})=\frac{1}{4\pi^{2}}\mbox{e}^{-|\alpha-(\alpha^{+})^{\ast}|^{2}/4}\Big\langle\frac{\alpha+(\alpha^{+})^{\ast}}{2}\Big|\rho
\Big|\frac{\alpha+(\alpha^{+})^{\ast}}{2}\Big\rangle,
\label{eq:Pdef}
\end{equation}
which we wish to sample for a Fock state with $\rho=|n\rangle\langle n|$. Introducing the new variables
\begin{equation}
\mu=\frac{\alpha-(\alpha^{+})^{\ast}}{2}\:\:\mbox{and}\:\:\gamma=\frac{\alpha+(\alpha^{+})^{\ast}}{2},
\label{eq:newvars}
\end{equation}
we find the separable expression
\begin{eqnarray}
P(\mu,\gamma) &=& \frac{\mbox{e}^{-|\mu|^{2}}}{\pi}\frac{|\gamma|^{2n}\mbox{e}^{-|\alpha|^{2}}}{\pi n!}\nonumber\\
&=& \frac{\mbox{e}^{-|\mu|^{2}}}{\pi}\frac{\Gamma(|\gamma|^{2},n+1)}{\pi},
\label{eq:gamma}
\end{eqnarray}
where 
\begin{equation}
\Gamma(x,n) = \frac{\mbox{e}^{-x}x^{n-1}}{(n-1)!}
\label{eq:gammadef}
\end{equation}
is the Gamma distribution. The variable $\mu$ is easily sampled via standard methods, while the Gamma distribution is sampled using a method given by Marsaglia and Tsang~\cite{Marsaglia} to give $z=|\gamma|^{2}$, so that $\gamma=\sqrt{z}\;\mbox{e}^{i\theta}$, where $\theta$ is uniform on $[0,2\pi)$. We then invert to find
\begin{equation}
\alpha = \mu+\gamma\:\:\mbox{and}\:\:\alpha^{+}=\gamma^{\ast}-\mu^{\ast},
\label{eq:inversion}
\end{equation}
which are now correctly distributed to represent the Fock state $|n\rangle$. For $n=1$, the decay follows the well-known exponential law, as asking whether a single isolated atom is bosonic or fermionic is a meaningless question. In fact, it is much simpler to calculate the decay for Fock states using a master equation formalism rather than stochastic equations~\cite{Stojan} and we present the result shown in Fig.~\ref{fig:Fock} more as evidence that our phase-space approach gives reliable results. We find that the stochastic results for larger $n$ agree with the master equation results wherever the integration converges. This gives us confidence that the results for other quantum states, for which a master equation solution is not so easily found, will also be accurate.

\begin{figure}
\begin{center} 
\includegraphics[width=0.9\columnwidth]{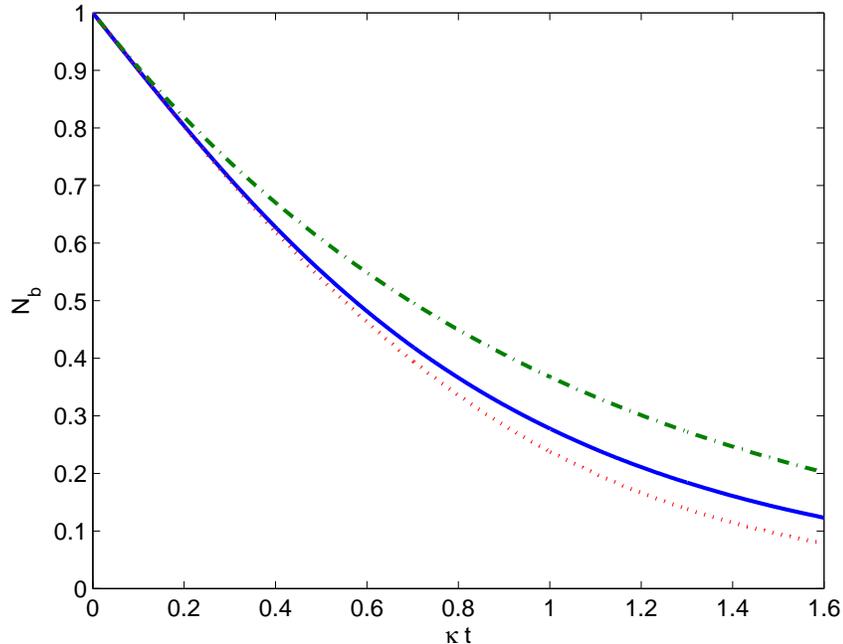}
\end{center} 
\caption{Stochastic result (solid line) from Eq.~\ref{eq:decaimento}, with $\kappa=0.2$ and $N_{b}(0)=1$, with the atom initially excited. For comparison, 
the standard exponential decay (dash-dotted line)
and the solution from Eq.~\ref{eq:Nbsoldet} (dotted line) are also shown. Note that in this result and that shown in Fig.~\ref{fig:NaNb}, 
the excited atom is initially in a coherent 
state. The equations were integrated over $9.9\times 10^{5}$ trajectories.}
\label{fig:stochdecay}
\end{figure}

\begin{figure}
\begin{center} 
\includegraphics[width=0.9\columnwidth]{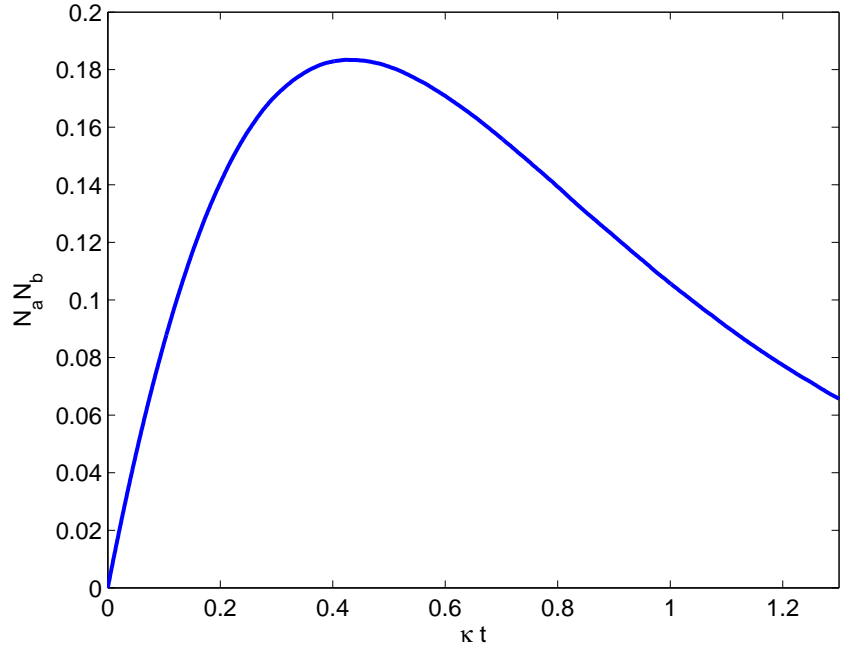}
\end{center} 
\caption{Stochastic solution for $\langle\hat{a}^{\dag}\hat{a}\hat{b}^{\dag}\hat{b}\rangle$, for the same parameters as used in Fig.~\ref{fig:stochdecay}.}
\label{fig:NaNb}
\end{figure}

In Fig.~\ref{fig:stochdecay} we show results for an initial coherent state, $|\beta\rangle$, where $\beta=1$ and, in Fig.~\ref{fig:NaNb}, the correlation which builds up between the ground and excited state populations for this initial state. For the initial $n=1$ Fock state, this correlation was zero. It can be seen that the decays are noticeably different, although we again note that as soon as we have the possibility of more than one atom being present, collisions and dipole-dipole interactions would also play a role. These have not been included in our analysis at this stage as we are more interested in a proof of principle rather than modelling exactly a given physical system. 

\begin{figure}
\begin{center} 
\includegraphics[width=0.9\columnwidth]{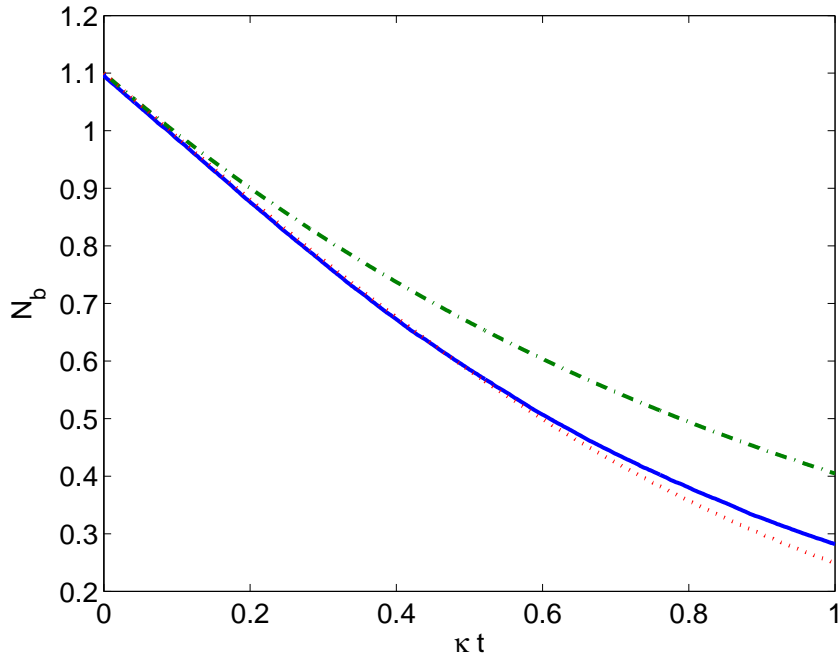}
\end{center} 
\caption{Decay of an initially excited ensemble with $1$ atom in the coherent component and an average of $0.1$ in the chaotic component. The stochastic average 
is shown as the solid line, while the 
analytical solution of Eq.~\ref{eq:Nbsoldet} is shown for comparison (dotted line), along with an 
exponential solutions with decay rate $\kappa$ (dash-dotted line).}
\label{fig:kasevich}
\end{figure}

\begin{figure}
\begin{center} 
\includegraphics[width=0.9\columnwidth]{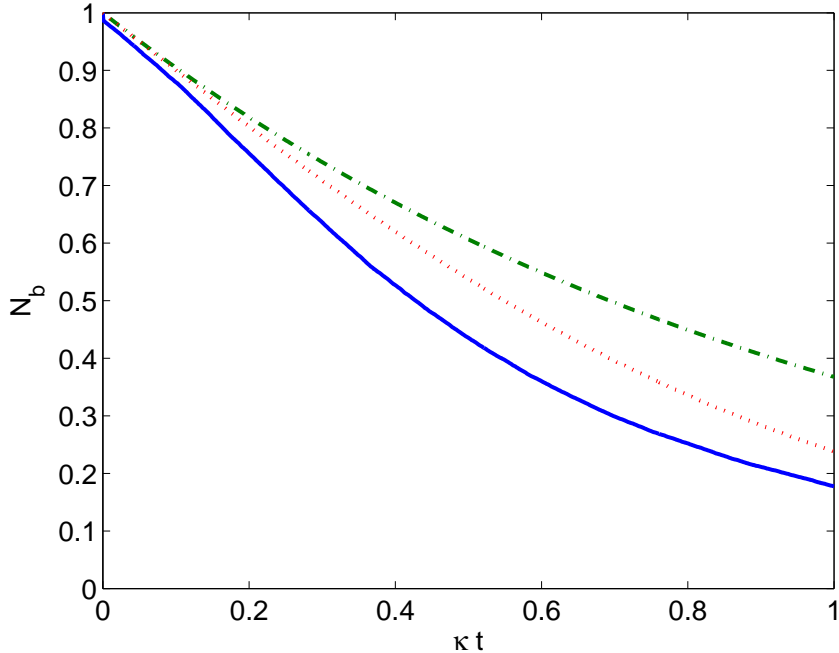}
\end{center} 
\caption{Decay of an initially excited ensemble with $\overline{n}=1$ and $\beta_{0}=0$. The stochastic average 
is shown as the solid line, while the 
analytical solution of Eq.~\ref{eq:Nbsoldet} is shown for comparison (dotted line), along with an 
exponential solutions with decay rate $\kappa$ (dash-dotted line).}
\label{fig:kasevich3}
\end{figure}

An interesting theoretical application of this model can be made for bosons in lattice wells. The Heisenberg uncertainty principle 
has been used to infer that atoms condensed at lattice sites were in number states, due to an increase in phase noise~\cite{Kasevich}. We note here that number states 
are not the only states with phase noise above the coherent state level, as chaotic and thermal states, among others, will also exhibit this property.
As small numbers of atoms in a lattice site may be a perfect candidate for measurements of the spontaneous emission rate, we will investigate this 
process for a chaotic state. These states have a particularly simple P-function, with
\begin{equation}
P(\beta)=\frac{1}{\pi\overline{n}}\exp(-|\beta|^{2}/\overline{n}),
\label{eq:Pchaos}
\end{equation}
where $\overline{n}$ is the average number present in the mode~\cite{Danbook}. If the state is a mixture of coherent and chaotic states, i.e. a chaotic state 
with a coherent displacement, the P-function is written as
\begin{equation}
P(\beta)=\frac{1}{\pi\overline{n}}\exp(-|\beta-\beta_{0}|^{2}/\overline{n}),
\label{eq:mix}
\end{equation}
where $\beta_{0}$ is the coherent displacement. We show the results of an averaging of $1.56\times 10^{5}$ stochastic trajectories in Fig.~\ref{fig:kasevich}, for $\beta_{0}=1$ and $\overline{n}=0.1$. 
In Fig.~\ref{fig:kasevich3}, we show the 
results for $\beta_{0}=0$ and $\overline{n}=1$. By comparison with Fig.~\ref{fig:stochdecay}, we see that the decay in the chaotic case is noticeably faster.

\section{Wigner representation} 
\label{sec:SED}

In the theory of stochastic electrodynamics~\cite{Trevor}, spontaneous emission is often claimed as being explicable as stimulated emission which is 
actually stimulated by vacuum fluctuations~\cite{Humberto}. The Wigner representation commonly used in quantum optics is equivalent to this theory if the 
Fokker-Planck equation for the Wigner function is truncated at second order, a positive-definite diffusion matrix is found, and distributions with positive Wigner functions are considered. 
Using the standard correspondences for It\^o calculus~\cite{Crispin}, the master equation (Eq.~\ref{eq:mestra}) can be mapped onto a generalised Fokker-Planck equation
for the Wigner function,
\begin{eqnarray}
\frac{dW}{dt} &=& \frac{\kappa}{2}\left\{-\left[\frac{\partial}{\partial\alpha}\left((|\beta|^{2}-\frac{1}{2})\alpha\right)+
\frac{\partial}{\partial\alpha^{\ast}}\left((|\beta|^{2}-\frac{1}{2})\alpha^{\ast}\right)\right.\right.\nonumber\\
& &\left.\left.
+\frac{\partial}{\partial\beta}\left(-(|\alpha|^{2}+\frac{1}{2})\beta\right)+\frac{\partial}{\partial\beta^{\ast}}
\left(-(|\alpha|^{2}+\frac{1}{2})\beta^{\ast}\right)\right]\right.\nonumber\\
& &\left.
+\frac{1}{2}\left[\frac{\partial^{2}}{\partial\alpha\partial\alpha^{\ast}}\left(2|\beta|^{2}-1\right)+\frac{\partial^{2}}{\partial\beta\partial\beta^{\ast}}\left(2|\alpha|^{2}+1\right)\right.\right.\nonumber\\
& &
\left.\left.
+\frac{\partial^{2}}{\partial\alpha\partial\beta}\left(-2\alpha\beta\right)+\frac{\partial^{2}}{\partial\alpha^{\ast}
\partial\beta^{\ast}}\left(-2\alpha^{\ast}\beta^{\ast}\right)
\right]\right.\nonumber\\
& & 
\left.
-\frac{1}{6}\left[\frac{\partial^{3}}{\partial\alpha\partial\alpha^{\ast}\partial\beta}\left(-\frac{3}{2}\beta\right)+
\frac{\partial^{3}}{\partial\alpha\partial\alpha^{\ast}\partial\beta^{\ast}}\left(-\frac{3}{2}\beta^{\ast}\right)\right.\right.\nonumber\\
& &
\left.\left.
+
\frac{\partial^{3}}{\partial\alpha\partial\beta\partial\beta^{\ast}}\left(\frac{3}{2}\alpha\right)+
\frac{\partial^{3}}{\partial\alpha^{\ast}\partial\beta\partial\beta^{\ast}}\left(\frac{3}{2}\alpha^{\ast}\right)
\right]
\right\}W(\alpha,\beta,t).
\label{eq:genWig}
\end{eqnarray}
Although methods exist for developing stochastic difference equations for a system which gives derivatives of higher than second-order in the 
Fokker-Planck equation~\cite{ourEPL,BWO}, we will truncate the above equation at second order as our aim is to compare the predictions of the positive-P representation with that 
of the representation which results from discarding the third-order derivatives in Eq.~\ref{eq:genWig}. However, what we do notice is that the 
diffusion matrix of Eq.~\ref{eq:genWig} is not positive-definite and thus has no straightforward mapping onto stochastic differential equations. As shown in Ref.~\cite{BWO}, we may double the phase-space in a way analogous to that used in the positive-P representation and find stochastic equations for 
four independent variables in what we may call a truncated positive-Wigner representation. 

We therefore make the changes $\alpha^{\ast}\rightarrow\alpha^{+}$ and $\beta^{\ast}\rightarrow\beta^{+}$, noting that averages of these are equivalent to symmetrically ordered operator expectation values, so that these are not the same as the variables used in the positive-P equations, although this should be obvious by context.
One possible factorisation of the diffusion 
matrix of Eq.~\ref{eq:genWig} is then
\begin{equation}
B_{W}= \left[\tilde{A}\:\tilde{0}_{4}\right]+\left[\tilde{0}_{4}\:\tilde{C}\right],
\end{equation}
where
\begin{equation}
\tilde{A} = \left[\begin{array}{cccc}
\frac{1}{2}\sqrt{\kappa(\beta^{+}\beta-\frac{1}{2})} & \frac{i}{2}\sqrt{\kappa(\beta^{+}\beta-\frac{1}{2})} & 0 & 0 \\
\frac{1}{2}\sqrt{\kappa(\beta^{+}\beta-\frac{1}{2})} & \frac{-i}{2}\sqrt{\kappa(\beta^{+}\beta-\frac{1}{2})} & 0 & 0 \\
0 & 0 & \frac{1}{2}\sqrt{\kappa(\alpha^{+}\alpha+\frac{1}{2})} & \frac{i}{2}\sqrt{\kappa(\alpha^{+}\alpha+\frac{1}{2})} \\
0 & 0 & \frac{1}{2}\sqrt{\kappa(\alpha^{+}\alpha+\frac{1}{2})} & \frac{-i}{2}\sqrt{\kappa(\alpha^{+}\alpha+\frac{1}{2})} 
\end{array}\right],
\label{eq:SEDmatrixA}
\end{equation}
and 
\begin{equation}
\tilde{C} = \left[\begin{array}{cccc}
\frac{i}{2}\sqrt{\kappa\alpha\beta} & 0 & 
-\frac{1}{2}\sqrt{\kappa\alpha\beta} & 0 \\
0 & \frac{i}{2}\sqrt{\kappa\alpha^{+}\beta^{+}} & 0 &
-\frac{1}{2}\sqrt{\kappa\alpha^{+}\beta^{+}} \\
\frac{i}{2}\sqrt{\kappa\alpha\beta} & 0 & 
\frac{1}{2}\sqrt{\kappa\alpha\beta} & 0 \\
0 & \frac{i}{2}\sqrt{\kappa\alpha^{+}\beta^{+}} & 0 & 
\frac{1}{2}\sqrt{\kappa\alpha^{+}\beta^{+}}
\end{array}\right],
\label{eq:SEDmatrixC}
\end{equation}
and $\tilde{0}_{4}$ is the $4\times 4$ null matrix.
This allows us to write the 
following stochastic differential equations,
\begin{eqnarray}
\frac{d\alpha}{dt} &=& \frac{\kappa}{2}\left(\beta^{+}\beta-\frac{1}{2}\right)\alpha+\frac{1}{2}\sqrt{\kappa(\beta^{+}\beta-\frac{1}{2})}\left(\eta_{1}+i\eta_{2}\right)
-\frac{1}{2}\sqrt{\kappa\alpha\beta}\left(\eta_{7}-i\eta_{5}\right),\nonumber\\
\frac{d\alpha^{+}}{dt} &=& \frac{\kappa}{2}\left(\beta^{+}\beta-\frac{1}{2}\right)\alpha^{+}+\frac{1}{2}\sqrt{\kappa(\beta^{+}\beta-\frac{1}{2})}\left(\eta_{1}-i\eta_{2}\right)
-\frac{1}{2}\sqrt{\kappa\alpha^{+}\beta^{+}}\left(\eta_{8}-i\eta_{6}\right),\nonumber\\
\frac{d\beta}{dt} &=& -\frac{\kappa}{2}\left(\alpha^{+}\alpha-\frac{1}{2}\right)\beta+\frac{1}{2}\sqrt{\kappa(\alpha^{+}\alpha+\frac{1}{2})}\left(\eta_{3}+i\eta_{4}\right)
+\frac{1}{2}\sqrt{\kappa\alpha\beta}\left(\eta_{7}+i\eta_{5}\right),\nonumber\\
\frac{d\beta^{+}}{dt} &=& -\frac{\kappa}{2}\left(\alpha^{+}\alpha-\frac{1}{2}\right)\beta^{+}+\frac{1}{2}\sqrt{\kappa(\alpha^{+}\alpha+\frac{1}{2})}\left(\eta_{3}-i\eta_{4}\right)
+\frac{1}{2}\sqrt{\kappa\alpha^{+}\beta^{+}}\left(\eta_{8}+i\eta_{6}\right).\nonumber\\
\label{eq:WigSDE}
\end{eqnarray}
These can again be readily integrated numerically, at least for states which have a well-behaved Wigner function. 

\begin{figure}
\begin{center} 
\includegraphics[width=0.9\columnwidth]{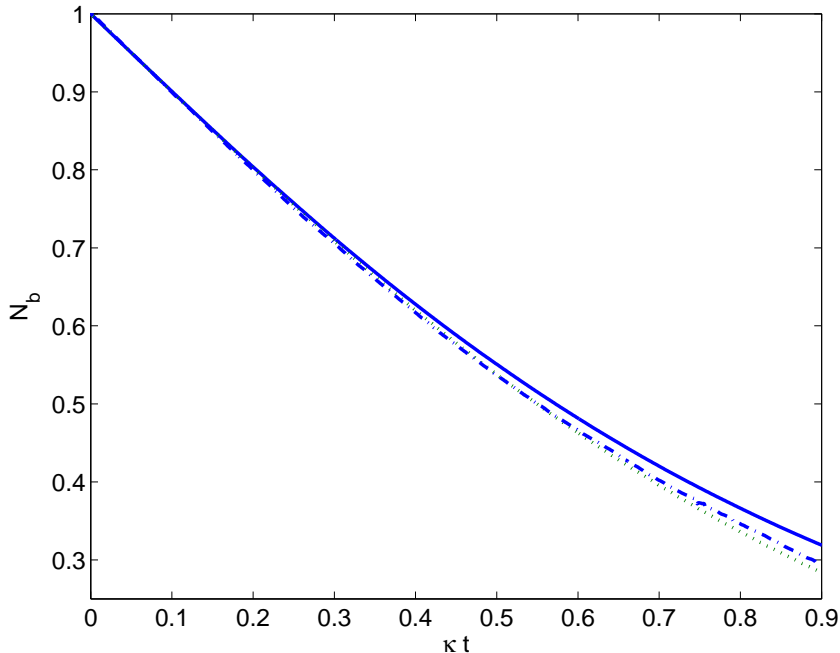}
\end{center} 
\caption{Excited state probability for an initially excited one-atom coherent state. The solid line is the positive-P average of $9.9\times10^{5}$ trajectories, the dash-dotted line represents
$3\times 10^{5}$ trajectories of Eq.~\ref{eq:WigSDE}, and the dotted line is the solution given in Eq.~\ref{eq:Nbsoldet}.}
\label{fig:Wplus}
\end{figure} 

We present a numerical solution to this equation, for an initial coherent state with an average of one excited atom, in Fig.~\ref{fig:Wplus}, along with the positive-P solution and the mean-field solution of Eq.~\ref{eq:Nbsoldet}. What we see is that, even though the phase-space has been doubled so that we are representing nonclassical dynamics, the truncated positive Wigner solution is closer to the mean-field prediction than it is to the full quantum solution. If the naive procedure of dropping the noise terms completely from Eq.~\ref{eq:WigSDE} and integrating the resulting equations is followed, the solutions do not even conserve atom number. This clearly suggests that any explanation of spontaneous emission as being due to vacuum fluctuations will not result in an accurate description of the dynamics. This is a demonstration of the failure of the truncated Wigner method for a system which is noticeably simpler than some of those for which it has previously been shown to give misleading results~\cite{asc,claus}. 

\section{Conclusions}

Using the example of spontaneous emission into a zero temperature reservoir, we have shown how stochastic equations may be developed to model the interaction of bosonic atoms with the electromagnetic field. This approach allows for the straightforward inclusion of the atomic quantum states, different numbers of atoms, the spatial dependence of atomic ensembles and interactions between the atoms, most of which would be difficult in the usual master equation approaches. We have shown that, except in the very special case of a one-atom Fock state, the decay is not exponential, but depends on correlations between the levels as well as bosonic stimulation by the population of the lower level. This is manifestation of the many-body atom statistics, because the 
other possible reason, superradiation, is eliminated by our choice of the 
model Hamiltonian. Our approach can be extended to describe the dynamics of more complicated processes such as, for example, electromagnetically induced transparency in degenerate gases. In principle it can also be extended to degenerate fermionic atoms using phase-space methods which are under development.

\section{Acknowledgments}

This research was supported by the Australian Research Council, the Ministero dell'Istruzione, dell'Universit\`a e della Ricerca (PRIN 2003), and 
the Landesstiftung Baden-Wurttemberg (grant No 33854).


\begin{thebibliography}{99}

\bibitem{Einstein}{A. Einstein, Phys. Z {\bf 18}, (1917), 121.}
%
\bibitem{WW}{E.P. Wigner and V. Weisskopf, Z. Physik {\bf 63}, (1930), 54.}
%
\bibitem{Agarwal}{G.S. Agarwal, \emph{Quantum Statistical Theories of Spontaneous Emission and their Relation to Other Approaches}, (Springer, Berlin, 1974.)}
%
\bibitem{Scott}{A.S. Parkins and D.F. Walls, Phys. Rep. {\bf 303}, (1998), 2.}
%
\bibitem{P+}{P.D. Drummond and C.W. Gardiner, J. Phys. A {\bf 13}, (1980), 2353.}
%
\bibitem{Wigner}{E.P. Wigner, \pr {\bf 40}, (1932), 749.}
%
\bibitem{Joel}{J.F. Corney and P.D. Drummond, \prl {\bf 93}, (2004), 260401.}
%
\bibitem{JCoriginal}{E.T. Jaynes and F.W. Cummings, Proc. IEEE {\bf 51}, (1963), 89.}
%
\bibitem{Italianos}{R. Bonifacio and G. Preparata, \pra {\bf 2}, (1970), 336.}
%
\bibitem{Crispin}{C.W. Gardiner, \emph{Quantum Noise}, (Springer, Berlin,
1991).}
%
\bibitem{Danbook}{D.F. Walls and G.J. Milburn, \emph{Quantum Optics}, (Springer, Berlin, 1994.)}
%
\bibitem{Glauber}{R.J. Glauber, \pr {\bf 131}, (1963), 2766.}
%
\bibitem{Sud}{E.C.G. Sudarshan, \prl {\bf 10}, (1963), 277.}
%
\bibitem{Sakurai}{J.J. Sakurai, \emph{Modern Quantum Mechanics}, (Addison-Wesley, Redwood City, 1985.)}
%
\bibitem{Rehler}{N.E. Rehler and J.H. Eberly, \pra {\bf 3}, (1971), 1735.}
%
\bibitem{Marsaglia}{G. Marsaglia and W.W. Tsang, ACM T. Math. Software, {\bf 26}, (2000), 363.}
%
\bibitem{Stojan}{S. Rebi\'c and M.K. Olsen, in preparation.}
%
\bibitem{Kasevich}{C. Orzel, A.K. Tuchman, M.L. Fenselau, M. Yasuda, M.A. Kasevich,
Science {\bf 291}, (2001), 2386.}
%
\bibitem{Trevor}{T.W. Marshall, Proc. R. Soc. London, Ser. A {\bf 276}, (1963), 475.}
%
\bibitem{Humberto}{H.M. Fran\c{c}a, T.W. Marshall, and E. Santos, \pra {\bf 45}, (1992), 6436.}
%
\bibitem{ourEPL}{L.I. Plimak, M.K. Olsen, M. Fleischhauer, and M.J. Collett, \epl {\bf 56}, (2001), 372.}
%
\bibitem{BWO}{L.I. Plimak, M. Fleischhauer, M.K. Olsen, and M.J.Collett, cond-mat/0102483.}
%
\bibitem{asc}{M.K. Olsen, K. Dechoum, and L.I. Plimak, \oc {\bf 190}, (2001), 261.}
%
\bibitem{claus}{C. Lamprecht, M.K. Olsen, P.D. Drummond, and H. Ritsch, \pra {\bf 65}, (2002), 053813.}
%


%------------------------------------------------------------------------------------------------------------------
\end{thebibliography}
\end{document}